\documentclass[12pt,a4paper]{article}
\usepackage[latin1]{inputenc}
\usepackage{graphicx,amsmath,cite}
\usepackage{amsfonts}
\usepackage{amssymb}
\begin{document}
%%%%%%%%%%%%%%%%%%%%%%%%%%%%%%%%%%%%%%%%%%%%%%%%%%%%%%%%%%%%%%%%%%%%%%%

\section*{Spontaneous $P$-parity violation in dense baryon matter\footnote{Talk
given at 11th International Conference on Meson-Nucleon Physics and
the Structure of the Nucleon (MENU2007), September 10-14 (2007), Juelich, Germany.}}
 %          {Spontaneous $P$-parity Violation}{S.S. Afonin \it{et al.}}
%\vspace{-6 cm}\includegraphics[width=6 cm]{Header.eps}
%\bigskip\bigskip
%\vspace{4 cm}

%\addcontentsline{toc}{chapter}{{\it S.S. Afonin}} \label{authorStart}
%%%%%%%%%%%%%%%%%%%%%%%%%%%% NEW SWITCHES %%%%%%%%%%%%%%%%%%%%%%%%%%%%%%

%%%%%%%%%%%%%%%%
%%% The authors
%%%%%%%%%%%%%%%%
\begin{center}
\large
S.S. Afonin$^{\star}$$^,$\footnote{E-mail address:
afonin24@mail.ru},
A.A. Andrianov$^{\$}$, V.A. Andrianov$^{\star}$,
D. Espriu$^{\$}$
\end{center}

\begin{raggedright}

$^{\star}$St. Petersburg State University, St. Petersburg, Russia\\
$^{\$}$University of Barcelona, Barcelona, Spain

\end{raggedright}

%\begin{center}
%{\bf Talk given at 11th International Conference on
%Meson-Nucleon Physics and
%the Structure of the Nucleon (MENU2007), September 10-14 (2007), Juelich,
%Germany}
%\end{center}

%%%%%%%%%%%%%
%% If the abstract fits to the title page remove the newpage commans
%%%%%%%%%%%%%
%\newpage

%\begin{center}
%\textbf{Abstract}
%\end{center}
\begin{abstract}
We investigate possibilities for dynamical $P$-parity violation in
dense baryon matter in the framework of effective quark models.
Dynamical $P$-parity violation can appear in models with at
least two scalar and two pseudoscalar fields, where both scalar
fields are condensed at normal conditions. At special
configurations of coupling constants, one of pseudoscalar fields
can then also condense at some value of baryon density, the
phenomenon results in mixing of the scalar and pseudoscalar
physical degrees of freedom, hence, giving rise to $P$-parity
violation. We discuss the implications and possible experimental
signatures for $P$-parity violation in strong interactions in future
experiments with heavy-ion collisions.
\end{abstract}
%\vspace{0.5cm}
%\section{Introduction}

\underline{\it Introduction.}
Presently the issue of dense baryon matter is attracting a lot of
interest as long as some striking physical phenomena are expected
to occur in certain regimes, such as the phase transition to
chirally symmetric hadron matter. The message we would like to
convey is that before any phase transition the $P$-parity in cold
dense baryon matter could undergo spontaneous breaking. At
zero baryon density (chemical potential) this phenomenon is
precluded by the Vafa-Witten theorem~\cite{witten}. However, the
conditions under which this theorem was proven (positivity of the
measure of partition function in vector-like theories) do not hold
anymore at finite baryon density (see~\cite{ae} for further discussions).

We shall report some important results of ongoing work along this
line. In short, at certain value of quark chemical potential and
in the physical range of model parameters, the phenomenon of
spontaneous parity breaking (SPB) has been recently observed in extensions
of popular low-energy models of QCD, namely in a generalized
Nambu--Jona-Lasinio model~\cite{ava} (the so-called Quasilocal Quark Model (QQM)),
in extended chiral quark
model~\cite{aet}, and in a generalized sigma-model~\cite{ae}. In
all cases, the underlying mechanism turned out to be rather
similar, we are going to describe briefly the relevant general features.

%\section{General analysis}
\underline{\it General analysis.}
The possibility of SPB arises when two different scalar fields condense with
a relative phase between their v.e.v.'s.
Let us consider a model with two multiplets of scalar/pseudoscalar fields
\begin{equation}
H_j = \sigma_j {\bf I} + i \hat\pi_j,
\quad j = 1,2;\quad H_j H_j^\dagger = (\sigma^2_j + (\pi^a_j)^2 ) {\bf I},
\end{equation}
where $\hat\pi_j \equiv \pi^a_j \tau^a$ with $\tau^a$ being a set of Pauli matrices.
We shall deal with scalar systems globally symmetric in respect to $SU(2)_L \times SU(2)_R$
rotations working in the exact chiral limit.
We should think of these two chiral multiplets as representing the two lowest-lying
radial states for a given $J^{PC}$. The introduced degrees of
freedom possess all the necessary ingredients to study SPB.

The effective potential of the models considered has, in general, the
following form at zero quark chemical potential $\mu$
(after specifying the v.e.v. $\langle H_1\rangle = \langle\sigma_1\rangle$),
\begin{multline}
V_{\text{eff}}= \frac12 \text{tr}\left\{- \sum_{j,k=1}^2 H^\dagger_j
\Delta_{jk} H_k + \lambda_1 (H^\dagger_1 H_1)^2 +
\lambda_2 (H^\dagger_2 H_2)^2+ \lambda_3 H^\dagger_1 H_1 H^\dagger_2 H_2
\right.\\
+ \frac12 \lambda_4 (H^\dagger_1 H_2 H^\dagger_1 H_2 +
H^\dagger_2 H_1 H^\dagger_2 H_1) + \frac12 \lambda_5
(H^\dagger_1 H_2 + H^\dagger_2 H_1) H^\dagger_1 H_1\\
+ \left.\frac12 \lambda_6 (H^\dagger_1 H_2 + H^\dagger_2 H_1) H^\dagger_2 H_2  \right\}
+ {\cal O}\left(\frac{|H|^6}{\Lambda^2}\right),
\label{effpot1}
\end{multline}
with 9 real constants $\Delta_{jk}, \lambda_i$.
QCD bosonization rules indicate that $\Delta_{jk}\sim \lambda_i \sim N_c$.
The neglected terms will be suppressed by inverse power of the chiral symmetry
breaking scale $\Lambda \simeq 1.2$~GeV. If we assume the v.e.v. of $H_j$
to be of the order of the constituent mass $0.2 \div 0.3$~GeV,  it is reasonable
to neglect these terms.

To guess the typical values of couplings, it is instructive to
consider a specific model. Let us take the QQM~\cite{ava}
as an example. The relevant form of Lagrangian is defined as follows,
\begin{equation}
{\cal L}_{\text{QQM}}=\bar q (i /\!\!\!\partial) q +
%\frac{1}{4 N_f N_{c}\Lambda^{2}}
\sum_{k,l=1}^{2}a_{kl}
\left[\bar q f_k(s)q \,\bar q f_l(s) q
- \bar q f_k(s) \tau^a \gamma_5 q \,
\bar q f_l(s)\tau^a \gamma_5 q\right].
\end{equation}
Here $a_{kl}$ represents a symmetric matrix of real coupling
constants and $f_k(s)$, $s \equiv -\partial^2/\Lambda^2$ are the
polynomial form factors specifying the quasilocal (in momentum
space) interaction. The form factors are orthogonal on the unit
interval and the results of calculations do not depend on a concrete
choice of form factors in the large-log approximation. A convenient choice
is $f_{1}(s) =2-3s$, $f_{2}(s) = -\sqrt{3}s$. The values of
couplings $\lambda_i$ in Eq.~\eqref{effpot1} are then fixed for $i=2,\dots,6$:
$\lambda_2=\frac{9N_c}{32\pi^2}$, $\lambda_3=\frac{3N_c}{8\pi^2}$, $\lambda_4=\frac{3N_c}{16\pi^2}$,
$\lambda_5=-\frac{5\sqrt{3}N_c}{8\pi^2}$, $\lambda_6=\frac{\sqrt{3}N_c}{8\pi^2}$.

We shall assume that the scalars under consideration are  generated in the quark sector of QCD.
The baryon chemical potential is transmitted to the meson sector via a quark-meson coupling.
Without loss of generality we can assume that only the first field $H_1$ has local coupling to quarks;
this actually defines the chiral multiplet $H_1$. Thus finite density is transmitted to the
boson sector via
$\Delta {\cal L}= - (\bar q_R H_1 q_L +\bar q_L H_1^\dagger q_R)$,
where $q_{L,R}$ are assumed to be constituent quarks. Then the one-loop contribution to $V_{\text{eff}}$ is
\begin{multline}
\Delta V_{\text{eff}}(\mu) = \frac{{\cal N}}{2} \Theta(\mu-|H_1|)\left[\mu|H_1|^2\sqrt{\mu^2-|H_1|^2}
- \frac{2\mu}{3}(\mu^2-|H_1|^2)^{3/2}\right.\\
\left.- |H_1|^4\ln{\frac{\mu+\sqrt{\mu^2-|H_1|^2}}{|H_1|}}\right]
\times \left(1+O\left(\frac{\mu^2}{\Lambda^{2}};
\frac{|H_1|^2}{\Lambda^{2}}\right)\right),
\label{potmu}
\end{multline}
where ${\cal N} \equiv \frac{N_{c}N_f}{4\pi^2}$ and $\mu$ is the chemical potential.
The higher-order contributions of chiral expansion in $ 1/\Lambda^2$  are not considered.
This effective potential is normalized to reproduce the baryon density
for quark matter
$\rho_B = \frac{N_f}{3\pi^2} p_F^3$,
where the quark Fermi momentum is\\ $p_F = \sqrt{\mu^2-|\langle H_1\rangle |^2}$.
Normal nuclear density is $\rho_B \simeq 0.17$ fm$^{-3} \simeq (1.8$ fm$)^{-3}$
that corresponds to the average distance $1.8$ fm between nucleons in nuclear matter.

Our analysis of the mass-gap equations and mass spectrum based on
potential~\eqref{effpot1} supplemented with the in-medium contribution~\eqref{potmu}
resulted in a generic picture which is graphically displayed in Fig.~1 and Fig.~2.

\begin{figure}
\begin{center}
%\hspace{-7cm}
\vspace{-3cm}
\includegraphics[scale=0.7]{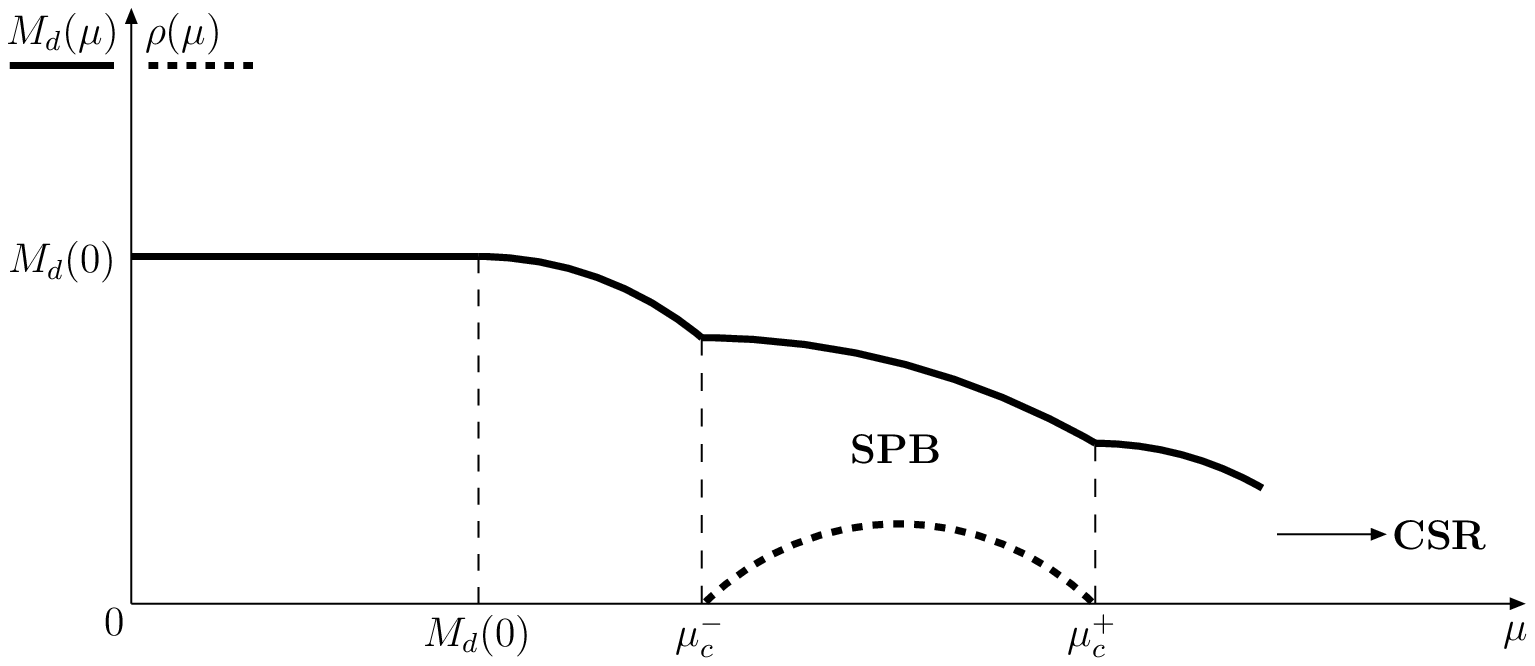}
\vspace{-13cm}
\caption{A qualitative dependence of dynamical quark mass $M_d$
and pseudoscalar condensate $\rho$ on quark chemical potential
$\mu$ (usually $M_d(0)\simeq300$~MeV). In the points of entering and
exiting the phase of spontaneous parity breaking (SPB) the derivatives
on $\mu$ jump. The region close to the chiral symmetry restoration (CSR)
is beyond the range of validity of chiral expansion.}
\label{F1}
%\vspace{-1cm}
\end{center}
\end{figure}
%\vspace{-1cm}
\begin{figure}
\begin{center}
\vspace{-3cm}
\includegraphics[scale=0.7]{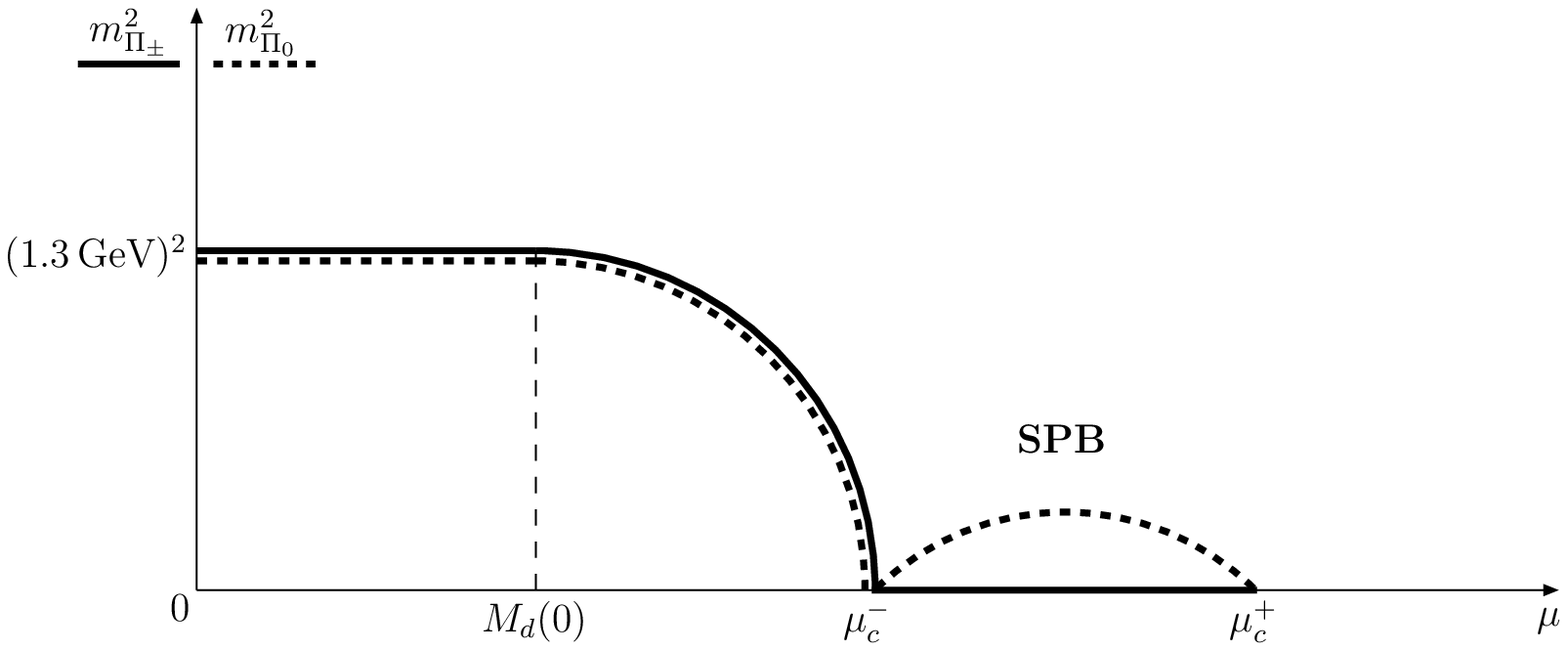}
\vspace{-13cm}
\caption{A qualitative behaviour of masses of isospin components for heavy
pseudoscalar meson as a function of quark chemical potential $\mu$ when
the SPB occurs.}
\label{F2}
\end{center}
\end{figure}

%\section{Discussions}
\underline{\it Discussions.}
Let us mention several possible signatures of $P$-parity breaking ensuing
from our analysis.

a) Decays of higher-mass meson resonances (radial excitations) into pions.
Resonances do not have a definite parity and therefore the same resonance can
decay both in two and three pions (in general into even and odd number of pions).

b) At the very point of the phase transition leading to parity breaking one has
six massless pion-like states. After crossing the phase transition, in the parity
broken phase, the massless charged pseudoscalar states remain as Goldstone bosons
enhancing charged pion production, whereas the additional neutral pseudoscalar state
becomes massive.

c) Reinforcement of long-range correlations in the pseudoscalar
channel (this effect could be hunted in lattice simulations).

d) Additional isospin breaking effects in the pion decay constant and substantial
modification of the weak decay constant $F_{\Pi^{\pm}}$ for massless charged excited pions,
giving an enhancement of electroweak decays.

%\section*{Acknowledgments}
\underline{\it Acknowledgments.}
This work is supported by research grants FPA2004-04582, SAB2005-0140 and RFBR 05-02-17477
as well as by Programs RNP 2.1.1.1112; LSS-5538.2006.2.
The work is also done with support of the EU RTN networks FLAVIANET and ENRAGE.

\end{document}